\documentclass[10pt, aps, prd, amsmath, amssymb, amsfonts, floats, floatfix, twocolumn, notitlepage,
superscriptaddress, nofootinbib, showpacs, longbibliography]{revtex4-2}
\usepackage{times}
\usepackage{xcolor}
\usepackage{bm}
\usepackage{dcolumn}
\usepackage{graphicx}
\usepackage{orcidlink}
\usepackage{hyperref}
\usepackage[utf8]{inputenc}
\usepackage[normalem]{ulem}
\usepackage{soul}
\usepackage[htt]{hyphenat}
\usepackage[caption=false]{subfig}
\usepackage{booktabs}
\usepackage{multirow}

\hypersetup{
    colorlinks=true,
    linkcolor=blue,
    citecolor=blue,
    urlcolor=blue,
    pdftitle={Biases in TGR from microlensed GW signals},
    }
\newcommand{\sigmaQGR}{\sigma_{\mathcal{Q}_{\rm GR}}}
\newcommand{\bayes}{\mathcal{B}^{\rm ML}_{\rm UL}}
\newcommand{\fml}{f_{\rm ML}}

\begin{document}

\title{Biases in Tests of General Relativity from Microlensed Gravitational-Wave Signals}

\author{Anirban Kopty\orcidlink{0009-0002-8933-9547}}
\email{anirban.kopty@iucaa.in}
\affiliation{Inter-University Centre for Astronomy and Astrophysics (IUCAA), Post Bag 4, Ganeshkhind, Pune 411007, India}

\author{Apratim Ganguly\orcidlink{0000-0001-7394-0755}}
\email{apratim@iucaa.in}
\affiliation{Inter-University Centre for Astronomy and Astrophysics (IUCAA), Post Bag 4, Ganeshkhind, Pune 411007, India}

\author{N. V. Krishnendu\orcidlink{0000-0002-3483-7517}}
\email{k.naderivarium@bham.ac.uk}
\affiliation{Institute for Gravitational Wave Astronomy \& School of Physics and Astronomy, University of Birmingham, Birmingham, B15 2TT, United Kingdom}

\author{Anuj Mishra\orcidlink{0000-0002-2580-2339}}
\email{anuj.mishra@icts.res.in}
\affiliation{International Centre for Theoretical Sciences, Tata Institute of Fundamental Research, Bengaluru - 560089, India}

\date{\today}

\begin{abstract}
    Gravitational-wave (GW) observations of compact binary mergers enable precision tests of general relativity (GR) in the strong-field regime, but their reliability depends on accurate waveform modeling. Unmodeled physical effects can introduce systematic biases that mimic apparent deviations from GR. In this work, we investigate the impact of GW microlensing on standard tests of GR employed by the LIGO--Virgo--KAGRA collaboration. Using a set of GW150914-like simulated signals lensed by isolated point-mass objects with masses in the range $10$--$10^{5} \, M_\odot$, we perform Bayesian parameter estimation with unlensed waveform templates and quantify biases in parameterized tests (TIGER and FTI), modified dispersion relation (MDR) tests, the inspiral--merger--ringdown consistency test (IMRCT), and the recently proposed meta-IMRCT framework.  To compare the outcomes of these tests using a common discriminator, we introduce a unified GR-deviation significance statistic based on the GR quantile that is applicable to both one- and multi-dimensional deviation parameters. We find that microlensing-induced waveform distortions can produce significant false deviations from GR, reaching ${\sim} 4\sigma$  in one-dimensional tests and ${\sim} 4.5\sigma$ in two-dimensional consistency tests, despite the injected signals being fully compatible with GR. The resulting false deviations arise predominantly in the wave-optics regime, where diffraction induces frequency-dependent amplitude and phase modulations, while simulated signals in the long-wavelength and geometric-optics regimes remain largely consistent with GR.  We further find that the significance of these apparent deviations does not correlate strongly with the detectability of microlensing as quantified by Bayes factors, indicating that GR tests probe specific projections of the waveform that are not captured by global lensing diagnostics. Our results establish gravitational lensing in the wave-optics regime as a potentially important astrophysical systematic for present and future precision tests of GR, highlighting the need to account for propagation effects in next-generation GW analyses.
\end{abstract}

\maketitle

\section{Introduction}
\label{sec:intro}

The advent of GW astronomy has opened a new observational window into the strong-field, highly dynamical regime of gravity, enabling unprecedented tests of GR~\cite{Arun:2013bp, Krishnendu:2021fga, Yunes:2025xwp}.
Following the detection of binary mergers reported by the LIGO--Virgo--KAGRA (LVK) collaboration~\cite{LIGOScientific:2018mvr, LIGOScientific:2020ibl, LIGOScientific:2021usb, KAGRA:2021vkt, LIGOScientific:2025slb, LIGOScientific:2026sit}, it is now possible to perform a comprehensive suite of tests using GW observations.
These include parameterized tests of waveform generation, consistency tests between different stages of the binary coalescence, tests of GW propagation, searches for residual signals and non-standard polarization states, and black hole (BH) ringdown analyses~\cite{LIGOScientific:2019fpa, LIGOScientific:2020tif, LIGOScientific:2021sio, LIGOScientific:2026qni, LIGOScientific:2026fcf, LIGOScientific:2026wpt}.
In these tests, theoretical waveform models are compared against the observed GW signals.
Any disagreement between the model and the data may therefore indicate a possible deviation from GR, modulo statistical uncertainties and non-Gaussian or non-stationary noise artifacts in the data.
Consequently, the robustness of these tests depends critically on the accuracy of the waveform models, which must incorporate all relevant physical effects within GR while maintaining systematic errors below the statistical uncertainties of the observations.
Any unmodeled physical effect that alters the observed waveform morphology can therefore introduce systematic biases that may be misinterpreted as evidence for deviations from GR~\cite{Gupta:2024gun}.
A growing body of recent work has shown that such effects--including orbital eccentricity~\cite{Saini:2022igm, Bhat:2022amc, Narayan:2023vhm, Shaikh:2024wyn}, waveform systematics~\cite{Moore:2021eok}, and gravitational lensing~\cite{Narayan:2024rat, Mishra:2023vzo}--can lead to apparent deviations from GR even when the true underlying theory is unmodified.
These false positives, or biases, have important implications for the reliability of GR tests, especially as detectors' sensitivity improves and we begin to probe more diverse source populations.

Gravitational lensing of GWs is one such propagation-induced waveform distortion that occurs when massive objects, such as galaxies, dark matter halos, or even stellar-mass BHs, bend spacetime enough to deflect and (de-) amplify GW signals~\cite{Einstein:1936llh, Zwicky:1937zzb}.
In the geometric-optics regime, lensing manifests as an overall amplification factor and possible multiple resolvable images.
Although lensing conserves the phase (except type-II images) and polarization structure in geometric optics, it modifies the apparent amplitude and arrival time of the signal.
More subtly, in certain lensing configurations, particularly those involving saddle or type-II images~\cite{Dai:2017huk}, the waveform morphology itself can deviate from unlensed expectations, making its detection possible~\cite{Ezquiaga:2020gdt, Janquart:2021nus, Vijaykumar:2022dlp}.
However, when the lens is a compact object with mass in the range $10$-$10^{5}\,\mathrm{M}_\odot$, such as stars, BHs, or compact dark matter clumps, the wavelength of GWs detected by ground-based interferometers becomes comparable to the lens's Schwarzschild radius.
In this \emph{wave-optics regime} (also known as \emph{microlensing})~\cite{Nakamura:1999uwi, Takahashi:2003ix}, the lensing effect no longer produces multiple resolvable images.
Instead, diffraction patterns are induced in the signal.
Such oscillatory distortions may influence the performance of waveform model-based GR tests considered in this work.

In this paper, we investigate how lensing in the wave-optics regime biases a representative set of GW tests of GR.
Specifically, we consider parameterized waveform-deviation tests (TIGER and FTI)~\cite{Agathos:2013upa, Mehta:2022pcn}, propagation tests based on modified dispersion relations (MDR)~\cite{Mirshekari:2011yq}, inspiral--merger--ringdown consistency tests (IMRCT)~\cite{Ghosh:2016qgn, Ghosh:2017gfp}, and the recently proposed meta-IMRCT framework~\cite{Madekar:2024zdj}.
These tests probe different physical aspects of the GW signal, ranging from modifications to the binary dynamics of waveform generation to propagation effects and internal consistency between different phases of the coalescence.
Together, they provide a useful laboratory for understanding how microlensing-induced waveform distortions can masquerade as signatures of beyond-GR physics.

To quantify these effects, we perform a systematic injection study using GW150914-like binary BH (BBH) signals microlensed by isolated point-mass lenses~\cite{Schneider:1992bmb} spanning the relevant lens parameter space.
The simulated signals are recovered using unlensed waveform models, allowing us to isolate the biases introduced solely by microlensing.
To facilitate a uniform comparison across tests with different numbers of deviation parameters, we introduce a common GR-deviation significance statistic based on the GR quantile, which can be naturally generalized to higher-dimensional parameter spaces.

This work complements earlier studies of biases induced by orbital eccentricity, Type-II strong lensing, and mililensing~\cite{Narayan:2023vhm, Narayan:2024rat, Liu:2024xxn}, while significantly extending our previous investigation of wave-optics lensing~\cite{Mishra:2023vzo}.
Our systematic comparison reveals a remarkably consistent picture across all the GR tests considered.
Significant apparent deviations from GR arise almost exclusively within the wave-optics regime, while simulated signals in the long-wavelength and geometric-optics regimes remain largely consistent with GR.
Depending on the test, these false deviations can reach ${\sim}4$--$4.5\sigma$ for individual simulated signals despite the injections being fully consistent with GR.
We further find that the detectability of microlensing, quantified through Bayesian model selection, does not strongly correlate with the magnitude of the induced GR-test biases.
This indicates that the waveform features responsible for apparent GR violations are distinct from those that maximize the evidence for the microlensed hypothesis.
Our results, therefore, identify wave-optics microlensing as a potentially important astrophysical systematic for current and future tests of GR.

The paper is organized as follows. In Sec.~\ref{sec:theory}, we summarize the microlensing formalism and the GR tests considered in this work.
Section~\ref{sec:methods} describes the simulated signals, Bayesian inference framework, and implementation of the various tests.
In Sec.~\ref{sec:results}, we present the results of the various tests of GR and identify the regions of the microlensing parameter space in which apparent deviations from GR arise.
We discuss the physical interpretation and broader implications of our findings in Sec.~\ref{sec:discussion}.
Appendix~\ref{app:fml300} explains the origin of the characteristic $\fml \simeq 300~{\rm Hz}$ transition observed for GW150914-like systems.

\section{Wave-optics Microlensing and GW-based Tests of GR}
\label{sec:theory}

In this section, we summarize the wave-optics description of GW lensing and briefly review the null tests of GR considered in this work.
Our goal is to establish the physical origin of the waveform distortions induced by microlensing, and to highlight why such distortions can bias tests that probe complementary aspects of the GW signal.

\subsection{Microlensing}
\label{subsec:microlensing}

When the wavelength of a GW becomes comparable to the characteristic scale of a gravitational lens, wave-optics (diffraction) effects become important, and the geometric-optics approximation is no longer valid.
In this wave-optics regime, lenses such as stars, stellar-mass BHs, or compact dark-matter structures can induce frequency-dependent distortions in the observed GW signal, a phenomenon commonly referred to as GW microlensing~\cite{Nakamura:1999uwi, Takahashi:2003ix}.
The dimensionless frequency parameter $\omega$ characterizes the transition between geometric and wave optics, and is defined as:
\begin{equation}\label{omega}
    \omega = \frac{8\pi G M_{\rm Lz} f}{c^3},
\end{equation}
where $M_{\rm Lz}$ is the redshifted lens mass and $f$ is the GW frequency.

For $\omega \gg 1$, the GW wavelength is much smaller than the Schwarzschild radius of the lens, and the geometric-optics approximation holds.
In this limit, the lensing effect can be modeled as a sum over distinct image contributions with associated magnifications and time delays~\cite{Schneider:1992bmb}.
In contrast, for $\omega \sim 1$, interference between the unresolved lensed images gives rise to diffraction, and the observed signal appears as a single waveform with frequency-dependent amplitude and phase modulations.
For an isolated point-mass lens model that we consider here, an analytical solution for the lensing amplification factor can be obtained~\cite{Takahashi:2003ix}:
\begin{align}
\label{amp-factor}
    F(\omega,y) &= \exp\left[\frac{\pi \omega}{4} + \frac{i \omega}{2}\left( \ln\frac{\omega}{2}-2\phi_m(y)\right)\right]\, \nonumber \\
    &\quad\times\Gamma\left(1 - \frac{i\omega}{2}\right) \, {}_1F_1\left(\frac{i\omega}{2}, 1; \frac{i\omega y^2}{2}\right),
\end{align}
where $y$ is the source-lens impact parameter normalized by the Einstein radius, and ${}_1F_1$ is the confluent hypergeometric function and $\phi_m(y)$ is the Fermat potential at the stationary point.
The lensing amplification factor modifies the unlensed frequency-domain waveform $\tilde{h}_U(f)$ as:
\begin{equation}
\tilde{h}_L(f) = F(\omega,y) \cdot \tilde{h}_U(f),
\end{equation}
with $F(\omega,y)$ capturing both frequency-dependent amplitude modulations and phase shifts, the latter act as effective dispersive distortions and therefore have the potential to bias GR-deviation parameters if lensing is neglected during waveform recovery.

It is useful to define a characteristic ``microlensing frequency'' $f_{\rm ML}$ by $\omega(f_{\rm ML})=1$~\cite{Mishra:2023ddt},
\begin{equation}
    f_{\rm ML} \equiv \frac{c^3}{8\pi G M_{Lz}}
    \approx 0.8\times10^3\,{\rm Hz}
    \left(\frac{10\,{\rm M}_\odot}{M_{Lz}}\right),
\end{equation}
which delineates the following regimes relevant for ground-based detectors:
\begin{itemize}
    \item \emph{Long-wavelength regime} ($f\ll f_{\rm  ML}$):
    the amplification factor approaches unity and lensing effects are strongly suppressed.

    \item \emph{wave-optics regime} ($f\sim f_{\rm ML}$):
    diffraction becomes important and the waveform acquires pronounced frequency-dependent amplitude and phase modulations.

    \item \emph{Geometric-optics regime} ($f\gg f_{\rm ML}$):
    $F(\omega,y)$ approaches the coherent sum of two image contributions with
    magnifications $\sqrt{|\mu_{\pm}|}$ and phase offsets $\exp(2\pi i f \Delta t)$.
\end{itemize}
The wave-optics regime is therefore where gravitational lensing can most significantly mimic the structures probed by the GR tests described in the next section.

Realistic astrophysical scenarios may involve microlens population in intervening galaxies or dark matter halos, leading to wave-optics modulations on top of geometric-optics amplification~\cite{Diego:2019lcd, Mishra:2021xzz}. Such configurations introduce additional astrophysical complexity that is not considered in this paper.
Here, we implement microlensing corrections using the isolated point mass lens approximation, given in Eq.~\eqref{amp-factor}, across the frequency spectrum of ground-based detectors.
The simulated lensed signals (\emph{injections}) are then used for GR tests with recovery templates that ignore the lensing corrections, allowing us to isolate the biases on parameter recoveries due to microlensing.

\subsection{Null tests of GR}
\label{subsec:gr-tests}

In this work, we consider the standard suite of theory-agnostic null tests of GR routinely employed in LVK analyses~\cite{LIGOScientific:2019fpa, LIGOScientific:2020tif, LIGOScientific:2021sio, LIGOScientific:2026qni, LIGOScientific:2026fcf}.
These tests probe complementary aspects of the GW signal, ranging from parameterized modifications of the waveform generation and propagation to consistency checks between different portions of the observed signal.

\paragraph*{\textbf{Parameterized Tests:}}
These tests of GR introduce phenomenological deviations into the GW waveform and constrain them directly from the data~\cite{Blanchet:1994ex, Blanchet:1994ez, Arun:2006hn, Arun:2006yw, Yunes:2009ke, Mishra:2010tp, Li:2011cg, Li:2011vx}.
In GR, all deviation coefficients vanish identically; therefore, any statistically significant non-zero measurement would indicate a departure from the GR prediction.
Two complementary implementations are commonly employed.

The first is the Test Infrastructure for General Relativity (TIGER) framework~\cite{Agathos:2013upa, Meidam:2017dgf, Roy:2025gzv}, in which deviations are introduced directly into the three distinct phases of an inspiral--merger--ringdown (IMR) phenomenological waveform model.
Denoting the collection of post-Newtonian waveform phase coefficients by $\varphi_i$, the deviation from GR is parameterized as
\begin{equation}
    \varphi_i\rightarrow (1+\delta\hat{\varphi_i})\varphi_i\,,
\end{equation}
where $\delta\hat{\varphi}=0$ corresponds to GR.
Depending on the coefficient being modified, the test probes deviations in the inspiral, merger, or ringdown parts of the signal.

The second implementation is the Flexible Theory-Independent (FTI) framework~\cite{Mehta:2022pcn}, which introduces generic deviations into the inspiral phase evolution of the frequency domain waveform model while remaining largely agnostic to the underlying model itself.
To avoid unphysical modifications near merger, the GR deviation coefficients are smoothly tapered to vanish at the inspiral to merger transition frequency.
Both TIGER and FTI therefore provide model-independent probes of potential departures from GR through modifications to the phase evolution of binary waveforms.

When constraining parameterized deviations, the additional parameters are varied one at a time rather than simultaneously.
This approach helps avoid parameter degeneracies and uninformative results that can arise when multiple parameters are varied simultaneously~\cite{Pai:2012mv}.
Although specific alternative theories may predict correlated modifications in multiple waveform coefficients, varying one deviation parameter at a time provides a robust and computationally efficient probe of generic departures from GR while avoiding the strong parameter degeneracies that arise in higher-dimensional inferences.

\paragraph*{\textbf{Modified Dispersion Tests:}}
GR predicts that GWs propagate non-dispersively in vacuum, implying that the speed of propagation is independent of the wave frequency, equal to the speed of light.
Their dispersion relation is $E^2=p^2c^2$, where $E$ and $p$ denote the energy and momentum of the graviton, respectively.
Many extensions of GR, including the theories with non-zero graviton mass, predict modified propagation laws leading to frequency-dependent GW velocities and hence dispersive propagation effects~\cite{Will:1997bb}.

A widely used phenomenological framework parametrizes such deviations through the modified dispersion relation~\cite{Mirshekari:2011yq, Samajdar:2017mka, LIGOScientific:2026fcf}
\begin{equation}
    E^2=\left(p c \right)^{2}+A_k\left(p c \right)^k,
\end{equation}
where $A_k$ is a phenomenological coefficient that characterizes the amplitude of the dispersive correction, while $k$ determines its frequency dependence.
Different values of $k$ can be mapped onto different classes of modified gravity scenarios and therefore provide a generic framework for testing a broad range of departures from GR.
For instance, the case $k=0$ corresponds to massive-graviton dispersion relation, with graviton mass $m_g=A_0^{1/2}c^{-2}$, for $A_0>0$.
In addition, depending upon the magnitude of $A_k$, one can classify the theories into superluminal $(A_k>0)$ and subluminal  $(A_k<0)$.

Modified GW propagation produces an accumulated frequency-dependent phase shift during propagation from the source to the detector.
In the frequency domain, the waveform can be written as
\begin{equation}
    \tilde h(f) = \tilde h_{\rm GR}(f) e^{i\delta\Psi(f)},
\end{equation}
where $\tilde h_{\rm GR}(f)$ denotes the GR waveform and $\delta\Psi(f)$ is the propagation-induced phase correction.
The explicit form of $\delta\Psi(f)$ depends on $k$, the source redshift, and the cosmological distance traveled by the GW.

\paragraph*{\textbf{Consistency Tests:}}
Unlike parameterized or propagation tests, consistency tests do not modify the waveform model itself.
Instead, they assess whether independent inferences drawn from the same GW signal are mutually consistent.

The IMR consistency test (IMRCT) compares the properties of the remnant BH inferred from the low-frequency (inspiral) and high-frequency (post-inspiral) parts of the signal independently~\cite{Ghosh:2017gfp}. Assuming GR describes the entire coalescence, the measured values of the final mass $M_f$ and dimensionless spin $\chi_f$ of the remnant BH from the two analyses should be consistent.
The fractional deviation parameters on the final mass and final spin are defined as
\begin{equation}
    \frac{\Delta M_f}{\bar{M}_f}=2\,\frac{M_f^{\rm insp}-M_{f}^{\rm postinsp}}{M_f^{\rm insp}+M_{f}^{\rm postinsp}},\,
    \label{eq:IMR-deviation-Mf}
\end{equation}
and
\begin{equation}
    \frac{\Delta \chi_f}{\bar{\chi}_f}=2\,\frac{\chi_f^{\rm insp}-\chi_{f}^{\rm postinsp}}{\chi_f^{\rm insp}+\chi_{f}^{\rm postinsp}}.
    \label{eq:IMR-deviation-chif}
\end{equation}
For a signal consistent with GR, both deviation parameters are expected to be centered on zero.
The degree of consistency is quantified by the GR quantile, $\mathcal{Q}_{\rm GR}$, defined as the fraction of the posterior probability enclosed by the isoprobability contour passing through the GR prediction $(\Delta M_f/\bar M_f,\Delta\chi_f/\bar\chi_f)=(0,0)$.

A recent extension of this framework is the meta-IMRCT~\cite{Madekar:2024zdj}.
Rather than comparing the inspiral and post-inspiral portions of a single analysis, the meta-IMRCT compares the remnant mass and spin inferred from different null tests of GR.
Since all GR tests, regardless of the specific physical modification it probes, should yield statistically consistent estimates of the remnant properties if GR is correct, discrepancies between them can serve as an additional indicator of departures from GR or of unmodeled waveform systematics.
The meta-IMRCT therefore provides a higher-level consistency check that combines information from multiple null tests and can reveal inconsistencies that may not be apparent in any individual test alone.

\section{Methodology}
\label{sec:methods}

In this section, we first describe the construction of the simulated BBH signals and the microlensing configurations considered in this work.
We then outline the parameter inference setup used to quantify biases in tests of GR, including the likelihood, priors, and sampling methodology, followed by a description of the individual GR tests employed in our analysis.

\subsection{Simulated signals and microlensed injections}
\label{subsec:injections}

To quantify the impact of microlensing on GW tests of GR, we perform a controlled injection study using simulated lensed BBH signals.
To isolate waveform systematics arising solely from microlensing, independent of statistical fluctuations due to detector noise, all analyses are performed using zero-noise injections.
All analyses assume a three-detector network consisting of the LIGO Hanford (H), LIGO Livingston (L) and Virgo (V) detectors~\cite{LIGOSensitivity, KAGRA:2013rdx, VIRGO:2014yos, Capote:2024rmo, LIGO:2024kkz} operating at their projected O5 design sensitivities\footnote{The PSDs were taken from~\url{https://dcc.ligo.org/ligo-t2000012/public}}.

As a representative source configuration, we consider GW150914-like BBH systems~\cite{gw150914}.
The intrinsic and extrinsic source parameters are fixed to the maximum-a-posteriori values reported in the GWTC-2.1 catalog~\cite{GWTC-2.1Zenodo}, including spin-precession.
The source-frame parameters correspond to a nearly equal-mass BBH with detector-frame total mass $M \simeq 72.7\,M_\odot$.
The luminosity distance $d_\mathrm{L}$ is rescaled for each injection such that the optimal network signal-to-noise ratio (SNR) is fixed to $30$.

Microlensing is modeled using the isolated point-mass lens model described in Sec.~\ref{subsec:microlensing}.
A total of 50 microlensed injections are generated on a logarithmically spaced grid in the lens parameters spanning
\begin{equation}
    M_{\mathrm{Lz}}/M_\odot \in [10^1,10^5],
    \qquad
    y \in [0.01,3],
\end{equation}
where $M_{\mathrm{Lz}}$ is the redshifted lens mass and $y$ is the dimensionless source position.
This parameter range uniformly covers the long-wavelength, wave-optics, and geometric-optics regimes relevant for ground-based GW detectors rather than representing an astrophysical population.
Configurations corresponding to the strong-lensing regime, for which the relative time delay between multiple images exceeds the chirp time of the GW signal such that the images would be observed as separate events, are excluded from the analysis.
Unless stated otherwise, unlensed injections are generated using the precessing IMR waveform approximant \texttt{IMRPhenomXPHM}~\cite{Pratten:2020ceb}.
Microlensing distortions are incorporated in the frequency domain through the amplification factor given in Eq.~(\ref{amp-factor}) using the publicly available package \textsc{gwmat}\footnote{\url{https://git.ligo.org/anuj.mishra/gwmat}}~\cite{Mishra:2023ddt}.

\subsection{Bayesian parameter inference framework}
\label{subsec:PE}

We perform Bayesian parameter estimation (PE) on the simulated GW signals (injections) to infer the source and microlensing parameters under both the unlensed and microlensed waveform hypotheses.
In Bayesian inference~\cite{Veitch:2014wba,sivia2006data,skilling2006nested}, the goal is to compute the posterior probability distribution, $p(\bm\theta|\bm d)$, such that $p(\bm\theta|\bm d) \, d\bm\theta$ represents the probability that the source parameters $\bm\theta$ lie within the infinitesimal parameter-space volume $(\bm\theta, \bm\theta+d\bm\theta)$ conditioned on the observed data $\bm d$.
Here, $\bm d$ denotes the detector strain data, while $\bm\theta$ collectively represents the model parameters to be inferred.
According to Bayes' theorem, the posterior distribution is given by,
\begin{equation}
    p(\bm\theta|\bm d) = \frac{\mathcal{L}(\bm d|\bm\theta) \, \pi(\bm\theta)}{\mathcal{Z}(\bm d)},
\end{equation}
where $\mathcal{L}(\bm d|\bm\theta)$ is the likelihood function, $\pi(\bm\theta)$ denotes the prior distribution on the parameters, and
\begin{equation}
    \mathcal{Z}(\bm d)\equiv \int d\bm\theta \,\mathcal{L}(\bm d|\bm\theta)\,\pi(\bm\theta)
\end{equation}
is a normalization factor, called Bayesian evidence.
For GW PE, we assume the standard stationary Gaussian-noise likelihood commonly employed in GW data analyses~\citep{Thrane:2018qnx, Veitch:2014wba}.
We adopt broad and astrophysically uninformative priors for all sampled parameters.

The Bayesian evidence, obtained by marginalizing the likelihood over the full parameter space, provides a natural framework for model selection between competing waveform hypotheses.
In particular, we compare two competing hypotheses:
\begin{itemize}
    \item $\mathcal{H}_{\rm UL}$: the GW signal is unlensed (null-hypothesis), and
    \item $\mathcal{H}_{\rm ML}$: the GW signal is microlensed.
\end{itemize}
The odds ratio between the two hypotheses is given by,
\begin{equation}
    \mathcal{O}^{\rm ML}_{\rm UL} = \frac{\mathcal{Z} (\bm d|\mathcal{H}_{\rm ML})}{\mathcal{Z} (\bm d|\mathcal{H}_{\rm UL})} \frac{\pi(\mathcal{H}_{\rm ML})}{\pi(\mathcal{H}_{\rm UL})}.
\end{equation}
Assuming equal prior probabilities for the two hypotheses, $\pi(\mathcal{H}_\text{UL}) = \pi(\mathcal{H}_\text{ML})$, the odds ratio reduces to the Bayes factor,
\begin{equation}
    \mathcal{O}^{\rm ML}_{\rm UL} = \bayes \equiv \frac{p(\mathcal{H}_{\rm ML}|\bm d)}{p(\mathcal{H}_{\rm UL}|\bm d)} = \frac{\mathcal{Z} (\bm d|\mathcal{H}_{\rm ML})}{\mathcal{Z} (\bm d|\mathcal{H}_{\rm UL})}.
    \label{eq:bayes-factor}
\end{equation}
The Bayes factor, therefore, quantifies the relative support provided by the data for the microlensed and unlensed waveform hypotheses.

All PE analyses are performed using the publicly available Bayesian inference library \textsc{Bilby}~\cite{Ashton:2018jfp, Romero-Shaw:2020owr}.
We employ the nested sampler \textsc{Dynesty}~\cite{Speagle:2019ivv}  with the \texttt{acceptance-walk} sampling method, using the following settings for each PE run: \texttt{nlive} $= 1000$, \texttt{naccept} $= 60$, and \texttt{n-parallel} $= 2$.
The simulated strain data are generated with a sampling rate of $2048~\mathrm{Hz}$ and analyzed using a minimum frequency cutoff of $20~\mathrm{Hz}$ for the likelihood evaluation, which is also the waveform reference frequency.
We employ both time and distance marginalization to accelerate likelihood evaluations; the former is justified by the negligible change in detector response over the time window of interest, while the latter introduces no additional assumptions.
Phase marginalization is not employed since our analysis involves higher-order multiple moments.

Table~\ref{tab:priors} summarizes the priors adopted for the intrinsic binary parameters, extrinsic parameters, and microlens parameters for all the PE runs.
For the GR analyses, the corresponding priors on the deviation parameters are listed in Table \ref{tab:TGR_priors}.
For a small subset of injections, the recovered posterior distributions exhibited railings at the boundaries of the default prior ranges.
These cases arise due to microlensing-induced biases pushing the recovered parameters beyond their expected range.
We reran using broader priors to ensure that the inferred posteriors were not artificially truncated by the prior boundaries.
\begin{table}[tb]
    {\renewcommand{\arraystretch}{1.2}%
    \begin{ruledtabular}
    \begin{tabular}{cc}
    Parameter & Prior Distribution \\ \hline \hline
     $\mathcal{M}~[M_\odot]$ & UniformCompMass%
     $(12, 120)$ \\
     $q$ & UniformCompMass%
     $(0.05, 1)$ \\
     $m_1,m_2~[M_\odot]$ & Constraint$(3, 500)$ \\
     $a_1,a_2$ & Uniform$(0, 0.99)$ \\
     $\theta_1,\theta_2$ & Sine$(0, \pi)$ \\
     $\phi_{12},\phi_{jl}$ & Uniform$(0, 2\pi)$ \\ \hline
     $\alpha$ & Uniform$(0, 2\pi)$ \\
     $\delta$ & Cosine$(-\pi/2, \pi/2)$ \\
     $\theta_{jn}$ & Sine$(0, \pi)$ \\
     $\psi$ & Uniform$(0, \pi)$ \\
     $\phi$ & Uniform$(0, 2\pi)$ \\
     $d_{\rm L}~[{\rm Mpc}]$ & UniformSource%
     $(500, 24000)$\\
     $t_c~[\rm s]$ & $t_{\rm merger} \pm 0.1$ \\ \hline
     $\log_{10}M_{\rm Lz}$ & Uniform$(-1, 5)$ \\
     $y$ & PowerLaw$(0.001, 5)$ with index $=1$ \\
    \end{tabular}
    \end{ruledtabular}
    }
    \caption{\label{tab:priors}%
    Prior distributions used for Bayesian PE. The parameters are grouped into intrinsic binary parameters, extrinsic parameters, and microlensing parameters. Under the microlensed hypothesis $\mathcal{H}_{\rm ML}$, all parameters are sampled simultaneously, while the lens parameters are excluded under the unlensed hypothesis $\mathcal{H}_{\rm UL}$. The \emph{UniformCompMass} prior corresponds to a distribution uniform in component masses, while \emph{UniformSource} denotes a prior uniform in comoving volume and source-frame time.}
\end{table}
%

\subsection{Fitting factor study}
\label{subsec:FF}

To complement the Bayes factor analysis, we also employ the fitting factor (FF)~\cite{Apostolatos:1995pj, Cornish:2011ys}, which quantifies how well one waveform family can recover signals generated by another.
We first define the normalized overlap between two waveforms, $h_t$ and $h$, as
\begin{equation}
\mathcal{O}(h_t,h) \equiv \frac{(h_t,h)}{\sqrt{(h_t,h_t)(h,h)}} = (\hat h_t,\hat h),
\end{equation}
where $(\cdot,\cdot)$ denotes the standard noise-weighted inner product, and $\hat h \equiv h/\sqrt{(h,h)}$ is the normalized waveform. The match is then defined as the overlap maximized over the coalescence time and phase,
\begin{equation}
\mathcal{M}(h_t,h) \equiv \max_{t_c,\phi_c}\,\mathcal{O}(h_t,h).
\end{equation}
Finally, the FF is obtained by maximizing the match over the intrinsic parameters $\theta$ of the template family,
\begin{equation}
\mathrm{FF} \equiv \max_{\theta}\,\mathcal{M}\!\left(h_t(\theta),h\right) = (\hat h_t^\text{max}, \hat h).
\end{equation}

Equivalently, FF can also be interpreted as the fraction of the true SNR, $\rho_\text{true} = \sqrt{(h,h)}$, recovered by the matched-filter SNR $\rho_\text{MF} = (\hat h_t^\text{max},h)$ using the best-matching template in the family, as in $\rho_\text{MF} = \text{FF} \times \rho_\text{true}$. In other words, it measures the reduction in detectability incurred when an inaccurate template family is used to represent the signal. In our analysis, the target waveform corresponds to a microlensed signal, whereas the template family consists of unlensed waveforms. The FF, therefore, provides a direct measure of the fraction of the SNR recovered by the unlensed model, with the associated loss given by $1-\mathrm{FF}$.

\subsection{Implementation of GR tests}
\label{subsec:tgr}

We now describe the implementation of the range of null tests of GR introduced in Sec.~\ref{subsec:gr-tests}.
Throughout this work, all GR tests are applied to simulated microlensed signals using waveform models that neglect lensing corrections, allowing us to quantify how microlensing-induced waveform distortions can manifest as false deviations from GR.
In each test, the inferred GR-deviation parameters are expected to be consistent with the GR prediction in the absence of waveform systematics.
Any statistically significant shift from the GR expectation, therefore, quantifies the bias induced by microlensing.
\begin{table}[b]
    {\renewcommand{\arraystretch}{1.2}%
    \begin{ruledtabular}
    \begin{tabular}{ccc}
     Test & Parameter & Prior \\
     \hline
     TIGER & $d\chi_0, d\chi_4, db_2, dc_2$ & Uniform$(-5,5)$ \\
     FTI & $d\chi_0, d\chi_4$ & Uniform$(-20,20)$ \\
     MDR & $A_\text{eff}$ & Uniform$(-3 \times 10^{-19}, 3 \times 10^{-19})$ \\
      & $k$ & $\{0,3,4\}$
    \end{tabular}
    \end{ruledtabular}
    }
    \caption{\label{tab:TGR_priors}%
    Priors used for the GR-deviation parameters in the different null tests.
    In each analysis, one testing parameter is sampled jointly with the standard binary parameters listed in Table~\ref{tab:priors}, resulting in a total of 16 sampled parameters per run.
    }
\end{table}

\paragraph*{\textbf{Parameterized Tests:}}
For each microlensed injection, a single deviation parameter is allowed to vary jointly with the standard binary parameters.
The priors adopted for the testing coefficients are listed in Table~\ref{tab:TGR_priors}.

The TIGER analyses are performed using the frequency-domain spin-precessing IMR waveform model \texttt{IMRPhenomXPHM}~\cite{Pratten:2020ceb}, including higher-order multipole moments~\cite{Roy:2025gzv}.
Although the TIGER framework allows a substantially larger set of deviation parameters~\cite{LIGOScientific:2026fcf}, exploring all available testing parameters for each of the 50 injections would be computationally expensive.
We restrict our analysis to four representative deviation parameters probing all stages of the coalescence: $d\chi_0$ and $d\chi_4$ in the inspiral regime, $db_2$ in the merger regime, and $dc_2$ in the ringdown regime.

The FTI tests are carried out using the effective one-body based model \texttt{SEOBNRv4\_ROM}, through the implementation described in~\cite{Mehta:2022pcn}.
We consider the inspiral coefficients $d\chi_0$ and $d\chi_4$, enabling a direct comparison with the corresponding TIGER tests in the inspiral regime.
Following the standard implementation~\cite{LIGOScientific:2026fcf}, the tapering frequency is chosen as
\begin{equation}
f_{\rm c}^{\rm PAR}=0.35\,f_{\rm peak}^{22},
\end{equation}
where $f_{\rm peak}^{22}$ denotes the GW frequency at the maximum amplitude of the dominant $(2,2)$-mode calculated using \texttt{SEOBNRv4\_ROM} waveform model.
The tapering width is fixed to one GW cycle.

To avoid waveform-model systematics from contaminating the inferred deviations, we employ the same waveform family for injection and recovery within a given framework.
In particular, the TIGER analyses use \texttt{IMRPhenomXPHM} consistently for both injection and recovery, whereas the FTI analyses use microlensed injections generated using \texttt{SEOBNRv4\_ROM}.

\paragraph*{\textbf{Modified Dispersion Tests:}}
The MDR analysis is carried out following the standard LVK implementation~\cite{LIGOScientific:2026fcf, Baka:2025drk} in order to determine whether microlensing-induced phase distortions can mimic non-GR propagation effects.
We consider three representative values of the dispersion exponent, $k \in \{0,\,3,\,4\}$.
For each choice of $k$, we perform full Bayesian inference including the effective dispersion amplitude parameter $A_{\rm eff}$, whose prior is given in Table~\ref{tab:TGR_priors}.
The parameter $A_{\rm eff}$ is related to the underlying MDR amplitude $A_k$ through a cosmology-dependent rescaling and is used to improve sampling efficiency in the implementation of the test~\cite{Mirshekari:2011yq}.
All MDR analyses are carried out using the \texttt{IMRPhenomXPHM} waveform model.

\paragraph*{\textbf{Consistency Tests:}}
In the IMRCT framework, the inspiral and post-inspiral segments are separated in the frequency domain at the frequency $f_{\mathrm{ISCO}}$, defined as the dominant-mode GW frequency at the innermost stable circular orbit (ISCO) of the remnant Kerr BH~\cite{Ghosh:2017gfp}.
For each injection, an initial GR PE run is carried out assuming the unlensed hypothesis.
The resulting maximum a posteriori (MAP) parameters are then used to compute $f_{\rm ISCO}$ and subsequently perform two similar independent PE runs, restricting to the inspiral and post-inspiral frequency ranges of the signal.
We employ the \texttt{IMRPhenomXPHM} waveform model.

From the resulting posterior samples of the two runs, we infer the remnant mass and spin using numerical-relativity calibrated fits~\cite{Hofmann:2016yih, Healy:2016lce, Jimenez-Forteza:2016oae}.
These estimates are then used to construct the fractional deviation parameters $\Delta M_f / \bar{M}_f$ and $\Delta \chi_f / \bar{\chi}_f$ defined in Eqs.~(\ref{eq:IMR-deviation-Mf})--(\ref{eq:IMR-deviation-chif}).
Since all analyses employ identical priors on the component masses and spins (Table~\ref{tab:priors}), no additional posterior reweighting is performed.
The GR quantile, $\mathcal{Q}_{\rm GR}$, computed using the \texttt{summarytgr} executable of the \textsc{PESummary} package~\cite{Hoy:2020vys}, forms the basis of the GR-deviation statistic introduced in Sec~\ref{sec:results}.

For each simulated signal, the meta-IMRCT combines the standard GR analysis together with the TIGER, MDR, and IMRCT analyses%
\footnote{
    Microlensed recoveries are not included as there will be no TGR bias if the physical effect is already accounted for.
},
but we exclude the FTI analyses since they employ a different waveform model.
This yields ten independent remnant inferences per simulated signal, yielding $^{10}C_{2} = 45$ distinct pairs (see Table~\ref{tab:PE-runs}).
For each pair of tests, we compute the consistency parameters $\{\Delta M_f / \bar{M}_f, \Delta \chi_f / \bar{\chi}_f\}$ and then evaluate the corresponding GR quantile.
The resulting set of $p$-values, defined as $p = 1 - \mathcal{Q}_{\rm GR}$, is then combined using Simes' method, a modified Bonferroni procedure that controls the overall false-discovery rate while preserving sensitivity to genuine inconsistencies~\cite{Madekar:2024zdj}%
\footnote{
    The reference also discusses Hommel's method, but we do not compute it, as the overall inference remains unchanged.
}.

All waveforms are generated using the \textsc{LALSimulation}~\cite{lalsuite} through the \textsc{Bilby} framework~\cite{Ashton:2018jfp}, while GR-test analyses are performed using \textsc{Bilby\_TGR}.
A summary of all analyses performed in this work is given in Table~\ref{tab:PE-runs}.
\begin{table}[htbp]
    {\renewcommand{\arraystretch}{1.25}%
    \begin{ruledtabular}
    \begin{tabular}{*3c}
     Analysis & Configuration & \#runs \\ \hline
     GR PE & $\mathcal{H}_{\rm UL},\,\mathcal{H}_{\rm ML}$ & $2\times 50$ \\
     TIGER & $d\chi_0,\,d\chi_4,\,d b_2,\,d c_2$ & $4\times 50$ \\
     FTI & $d\chi_0,\,d\chi_4$ & $2\times 50$ \\
     MDR & $k \in \{0,3,4\}$ & $3\times 50$ \\
     IMRCT & Inspiral, Post-inspiral & $2\times 50$
    \end{tabular}
    \end{ruledtabular}
    }
    \caption{\label{tab:PE-runs}%
    Summary of the analyses performed for each of the 50 microlensed injections. The standard PE analysis includes both the unlensed ($\mathcal{H}_{\rm UL}$) and microlensed ($\mathcal{H}_{\rm ML}$) hypotheses.
    For the parameterized tests, one deviation parameter is varied at a time.
    In total, $13\times50=650$ PE runs were performed.
    The meta-IMRCT subsequently combines the results of the 10 independent analyses per simulated signal: one standard PE analysis ($\mathcal{H}_{\rm UL}$), two IMRCT analyses (inspiral and post-inspiral), four TIGER analyses, and three MDR analyses, yielding ${}^{10}C_2=45$ pairwise consistency tests.
    The FTI analyses are excluded from the meta-IMRCT construction because they employ a different waveform approximant.
    }
\end{table}
%

\section{Results}
\label{sec:results}

We now present the results of the null tests of GR performed on the simulated microlensed signals described in Sec.~\ref{subsec:injections}.
In total, $650$ Bayesian PE analyses were carried out (see Table~\ref{tab:PE-runs}).
Throughout this section, background shading is used to indicate the three lensing regimes introduced in Sec.~\ref{subsec:microlensing}: long-wavelength (light purple), wave-optics (pale yellow), and geometric-optics (light blue-green).
Foreground colors and marker styles represent the quantities being plotted and therefore vary across figures.
For posterior and parameter-bias plots, marker colors indicate the lensing regime of the corresponding injection (red: long-wavelength, green: wave-optics, blue: geometric-optics).
In contrast, figures showing statistical measures
use continuous color scales to represent the corresponding numerical values.

\subsection{Microlensing detectability}
\label{subsec:detect}

\begin{figure}[htbp]
    \includegraphics[width=\linewidth]{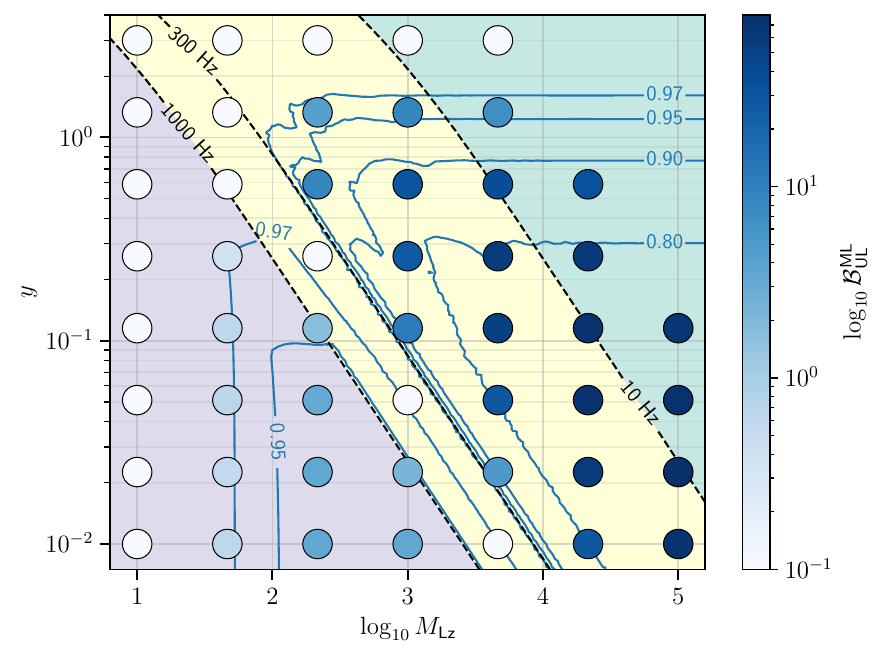}
    \caption{\label{fig:bayes}%
    Bayes factors shown for all 50 GW150914-like microlensed injections across the microlensing parameter space. The background shading indicates the three microlensing regimes, defined by $\fml$ at $10$ and $1000$ Hz: long-wavelength (light purple), wave-optics (pale yellow), and geometric-optics (light blue-green), spanning from the bottom left to the top right. Circles denote the Bayes factor values ($\log_{10} \bayes$). The geometric-optics regime also includes millilensing regime; configurations producing resolvable multiple signals are excluded, resulting in the empty region in the upper-right corner. Blue contours show the FF values obtained using unlensed waveform templates, a trend consistent with the Bayes-factor values. The line $\fml=300$ Hz approximately marks the transition between strongly and weakly detectable microlensing effects for the GW150914-like injections considered in this work.}
\end{figure}

Before examining biases in the null tests of GR, it is useful to quantify the detectability of microlensing itself.
Fig.~\ref{fig:bayes} summarizes the Bayes factor $\bayes$ between the microlensed ($\mathcal{H}_{\rm ML}$) and unlensed ($\mathcal{H}_{\rm UL}$) hypotheses across the microlens parameter space for the 50 GW150914-like microlensed injections.
For comparison, the figure also shows the FF contours obtained by maximizing the overlap between the microlensed injections and unlensed waveform templates.
Regions with low FF values correspond to injections whose diffraction-induced distortions cannot be efficiently absorbed by changes in the intrinsic unlensed binary parameters, resulting in larger mismatches.
Two clear trends emerge.
First, the Bayes-factor distribution closely follows the FF contours, confirming that the Bayesian model selection and mismatch-based approaches identify the same regions of parameter space as being most strongly affected by microlensing.
Second, the strongest support for the microlensed hypothesis is concentrated within the wave-optics regime, where diffraction-induced amplitude and phase modulations are most pronounced.
The region of significant support extends smoothly into the millilensing regime, where the signal approaches the coherent superposition of two lensed images separated by a finite time delay.
In addition, a narrow band of reduced microlensing support is visible in the long-wavelength regime.
Similar features have been reported in previous studies~\cite{Bondarescu:2022srx,Mishra:2023ddt,Chan:2024qmb}.

For the GW150914-like systems considered here, the Bayes factor decreases rapidly above $\fml \sim 300~{\rm Hz}$, indicating that microlensing-induced waveform distortions become progressively more difficult to distinguish from unlensed signals.
As demonstrated in Appendix~\ref{app:fml300}, this transition occurs because the characteristic diffraction-induced modulations move into frequency regions where the unlensed waveform carries relatively little signal power.
Consequently, although the formal wave-optics regime extends up to $\fml\sim1000~{\rm Hz}$, detectable microlensing signatures for GW150914-like binaries are largely confined to $\fml\lesssim300~{\rm Hz}$.
This characteristic scale will reappear throughout the following subsections, where we show that the largest apparent deviations from GR are similarly restricted to this region.

\subsection{Deviations from GR}
\label{subsec:gr-deviation}

We now investigate the extent to which microlensing-induced waveform distortions can be misinterpreted as apparent deviations from GR when analyzed with unlensed recovery model.
Before examining the microlensed injections, however, it is useful to establish a reference by performing the same analyses on an unlensed GW150914-like simulated signal.
All parameters are identical to those of the microlensed injections, except for the luminosity distance $d_{\mathrm{L}}$, which is rescaled to maintain an optimal network SNR of 30 for the HLV detectors at O5 sensitivity, matching the microlensed cases.
\begin{table*}[htb]
{\renewcommand{\arraystretch}{1.25}%
\begin{ruledtabular}
\begin{tabular}{cccccccccccc}
\multirow{2}{*}{$\log_{10}\bayes$}
 & \multicolumn{3}{c}{MDR}
 & \multicolumn{4}{c}{TIGER}
 & \multicolumn{2}{c}{FTI}
 & \multirow{2}{*}{IMRCT}
 & \multirow{2}{*}{meta-IMRCT} \\

\cmidrule(lr){2-4} \cmidrule(lr){5-8} \cmidrule(lr){9-10}

{} & $A_0$ & $A_3$ & $A_4$
   & $\delta\chi_0$ & $\delta\chi_4$ & $\delta b_2$ & $\delta c_2$
   & $\delta\chi_0$ & $\delta\chi_4$
   & {} & {} \\

\hline
$-0.17$ & $0.20$ & $0.05$ & $0.13$
        & $0.35$ & $0.03$ & $0.71$ & $0.16$
        & $0.52$ & $0.07$
        & $0.11$ & $0.08$ \\
\end{tabular}
\end{ruledtabular}
}
\caption{\label{tab:background}%
    Bayes factor and GR-deviation significances obtained for an unlensed GW150914-like injection. The quoted values correspond to $\sigmaQGR$ for the various null tests considered in this work.}
\end{table*}

To compare the different null tests on a common footing, we characterize the significance of an apparent GR violation through the GR quantile, $\mathcal{Q}_{\rm GR}$, introduced in Sec.~\ref{subsec:tgr}.
We convert this into an equivalent Gaussian significance, denoted by $\sigmaQGR$%
\footnote{
    The Mahalanobis distance is used as a distance measure in higher dimensions, which reduces to the normal distance in one dimension. One can understand the difference in the number of sigma deviations in different dimensions by the fact that, for the same distance, the volume contained by it increases with the increase in dimensions. This leads to a lower quantile value for the same sigma in higher dimensions.
}, according to
\begin{align}
    \sigmaQGR =
    \begin{cases}
        \sqrt{2} \, \text{erf}^{-1} (\mathcal{Q}_\text{GR}) \, & \text{for 1D},  \\[1ex]
        \sqrt{-2 \, \ln (1 - \mathcal{Q}_\text{GR})} \,  & \text{for 2D},
    \end{cases}
    \label{eq:sigma-QGR}
\end{align}
where the first expression applies to tests involving a single deviation parameter (TIGER, FTI, and MDR), while the second applies to the two-dimensional consistency tests (IMRCT and meta-IMRCT).
This definition provides a unified measure of apparent GR deviations across all tests considered in this work and can be straightforwardly generalized to higher-dimensional parameter spaces.
In practice, each PE run contains approximately $2\times10^4$ posterior samples, and finite sampling therefore limits the maximum measurable significance.
The largest resolvable deviation corresponds to approximately $4.05\sigma$ for one-dimensional tests and $4.45\sigma$ for two-dimensional tests.
Throughout this paper, we therefore quote significances above these values as approximately $4\sigma$ and $4.5\sigma$, respectively.

The results for the unlensed injection are summarized in Table~\ref{tab:background}.
As expected, the simulated signal does not favor the microlensed hypothesis, yielding a negative Bayes factor, $\log_{10}\bayes=-0.17$.
More importantly, all null tests remain consistent with the GR prediction.
The largest apparent deviation is only $0.71\sigmaQGR$, obtained for the TIGER parameter $\delta b_2$.
This demonstrates that the analysis pipelines themselves do not generate spurious GR violations and provides a reference against which the microlensed injections can be compared.

\begin{figure*}[t]
    \subfloat[\label{fig:MDR-posterior}]{%
        \includegraphics[width=0.53\textwidth]{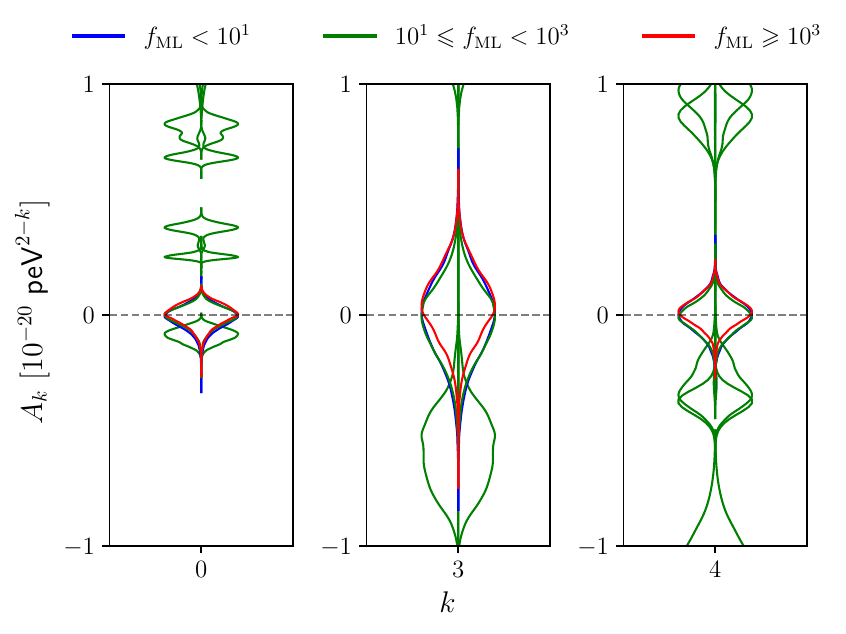}%
    }\hfill
    \subfloat[\label{fig:IMR-posterior}]{%
        \includegraphics[width=0.46\textwidth]{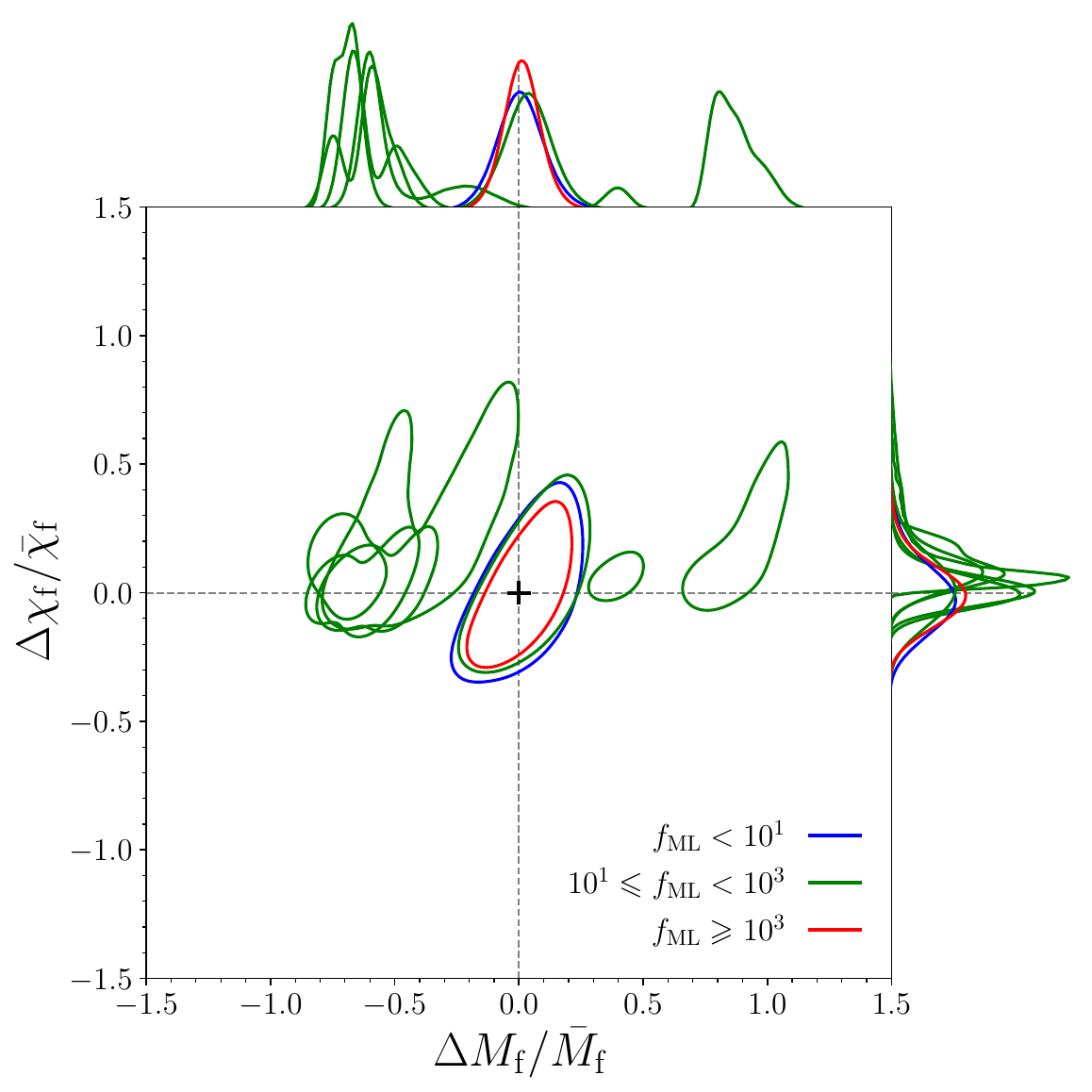}%
    }\hfill
    \caption{Posterior distributions shown for MDR parameter $A_k$, $k \in \{0,3,4\}$ as violin plots (left) and IMRCT deviation parameters $\{\Delta M_f/\bar M_f, \Delta \chi_f / \bar \chi_f\}$ as 95\% contours (right). In both panels, we display posteriors corresponding to simulated signals with at least a $3\sigmaQGR$ deviation. Representative posteriors are also shown for each of the three microlensing regimes: geometric-optics (blue), wave-optics (green), and long-wavelength (red). We observe that posteriors excluding the GR value predominantly arise from the wave-optics (green) regime, although not all of them exhibit significant deviations.}\label{fig:posteriors}
\end{figure*}
Having established the baseline behavior of the analyses pipelines, we now examine representative microlensed injections to illustrate how waveform distortions manifest as apparent deviations from GR.
Representative posterior distributions from the null tests are shown in Fig.~\ref{fig:posteriors}.
The left panel displays posterior distributions for the MDR parameters $A_k$ with $k\in\{0,3,4\}$, while the right panel shows the $95\%$ credible regions of the corresponding IMRCT posteriors in the $(\Delta M_f/\bar M_f,\Delta\chi_f/\bar\chi_f)$ plane.

Displaying results for all 50 injections would obscure the overall trends, since most injections in the long-wavelength and geometric-optics regimes yield posteriors centered on and enclosing the GR value.
We therefore focus only on those simulated signals exhibiting large apparent deviations from GR, along with one representative example from each of the three microlensing regimes.
A clear trend emerges: posterior distributions that exclude the GR prediction arise almost exclusively from injections belonging to the wave-optics regime.
At the same time, not every simulated signal in the wave-optics dominated regime exhibits a large deviation, indicating that microlensing may not be a sufficient condition for producing significant biases in GR tests.
For the IMRCT posteriors, the dominant contribution to the apparent inconsistency originates from the remnant-mass parameter $\Delta M_f/\bar M_f$, while the marginalized posteriors for $\Delta\chi_f/\bar\chi_f$ remain broadly centered on the GR value.

\paragraph*{\textbf{Parameterized Tests:}}

\begin{figure*}[p]
    \includegraphics[width=\linewidth]{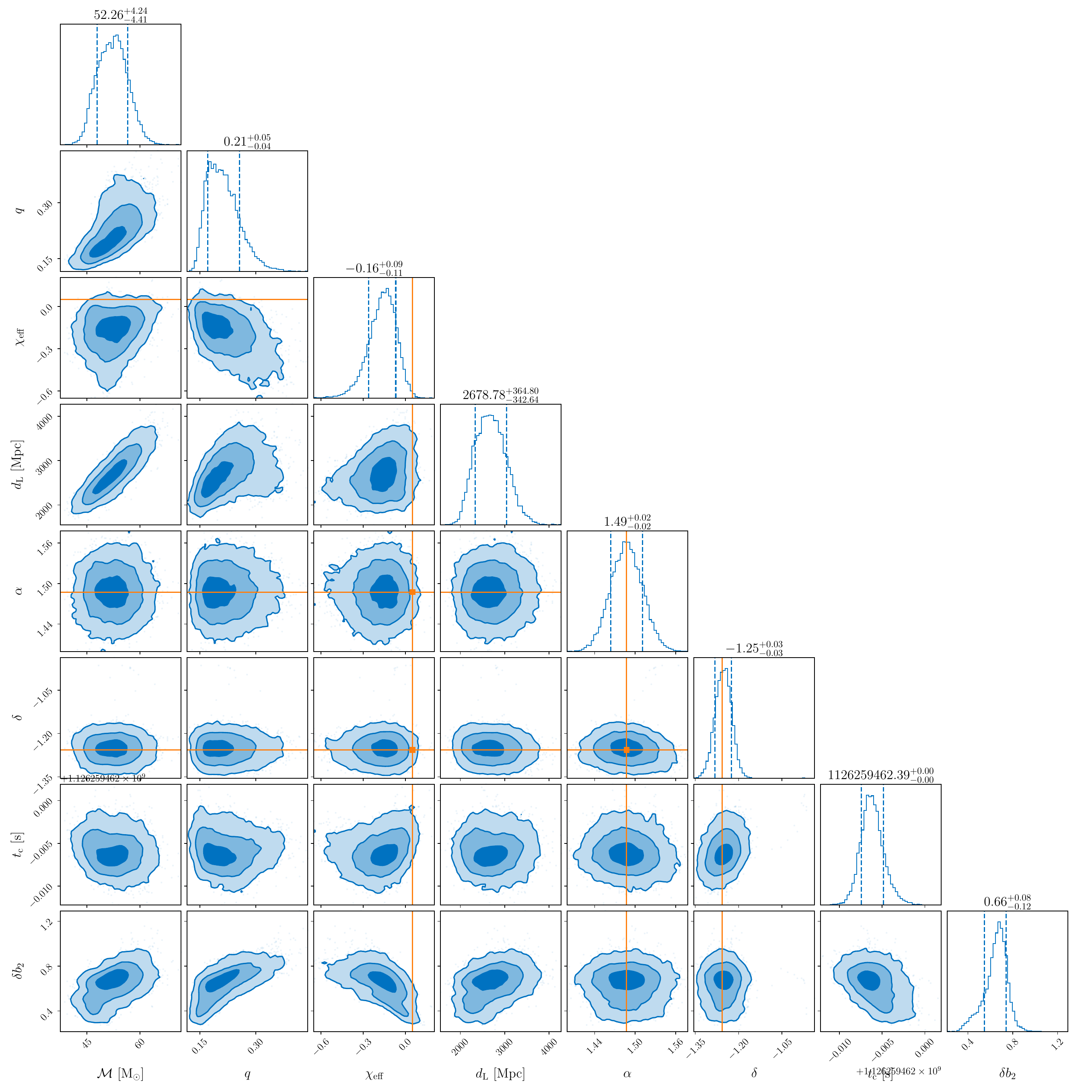}
    \caption{\label{fig:PE-posterior}%
    Posterior distributions from the TIGER analysis for the deviation parameter $\delta b_2$ for a GW150914-like microlensed signal with microlensing parameters $(\log_{10} M_{{\rm Lz}}, y) \approx (3, 0.115)$. The two-dimensional panels show $1\sigma$, $2\sigma$, and $3\sigma$ credible regions, while the one-dimensional histograms display the marginalized posteriors together with the median and $1\sigma$ credible intervals. The injected values are marked by orange lines; however, for several parameters, they lie outside the displayed posterior ranges, indicating strongly biased parameter recoveries. This wave-optics dominated simulated signal results in an apparent deviation from GR exceeding $4\sigmaQGR$.}
\end{figure*}
We now present the results obtained by performing TIGER and FTI analyses for all 50 microlensed injections.
As discussed in Sec.~\ref{subsec:tgr}, TIGER analyses employ the \texttt{IMRPhenomXPHM} waveform model, whereas FTI uses \texttt{SEOBNRv4\_ROM}.
In each framework, the same waveform approximant is used consistently for both injection and recovery to eliminate any systematic effects arising from waveform modeling.

Before examining the full injection set, it is instructive to consider a representative simulated signal exhibiting a large apparent deviation from GR.
Figure~\ref{fig:PE-posterior} shows the posterior distributions from a TIGER analysis of the deviation parameter $\delta b_2$ for a GW150914-like microlensed signal with $(\log_{10}M_{\rm Lz},y)\approx(3,0.115)$.
This simulated signal yields an apparent deviation exceeding $4\sigmaQGR$.
The injected source parameters lie outside the $3\sigma$ credible regions for most intrinsic and extrinsic binary parameters, demonstrating substantial biases in the recovered source properties.
Only the sky-location parameters $(\alpha, \delta)$ and, in this particular example, the coalescence time $t_c$ remain largely unaffected.
Such behavior is expected when an unlensed waveform model attempts to fit a microlensed signal.
Since the recovery model lacks the lensing degrees of freedom required to reproduce the diffractive modulations, the inference compensates by biasing the astrophysical source parameters as well as the GR-deviation parameter, leading to an apparent violation of GR.

\begin{figure}[htbp]
    \includegraphics[width=\linewidth]{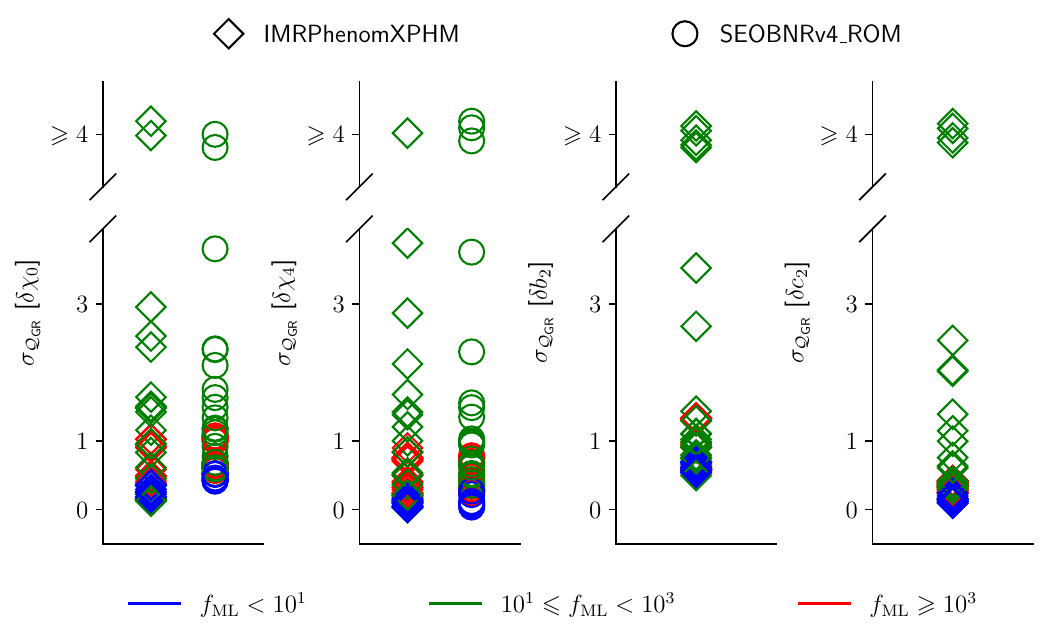}
    \caption{\label{fig:PAR}%
    Bias in parameterized test coefficients $\{\delta\chi_0, \delta\chi_4, \delta b_2, \delta c_2\}$ for GW150914-like microlensed injections. Each panel displays 1D marginalized posteriors for the deviation parameters, with diamond markers indicating the TIGER (\texttt{IMRPhenomXPHM}) results and circles representing the FTI (\texttt{SEOBNRv4\_ROM}) analyses. The y-axis represents the significance of the apparent GR deviation, quantified by $\sigmaQGR$. Marker colors denote the corresponding microlensing regimes characterized by $\fml$: geometric-optics (blue), wave-optics (green), and long-wavelength (red). The vertical spread of markers at $\sigmaQGR\geqslant 4$ is not quantitatively meaningful and serves only to separate overlapping simulated signals.
    We find that while some long-wavelength injections exceed $1\sigmaQGR$, all red and blue markers remain below $2\sigmaQGR$, whereas only wave-optics dominated injections exceed $3\sigmaQGR$.}
\end{figure}
The results for all 50 injections are summarized in Fig.~\ref{fig:PAR}.
The four panels show the inferred significances, $\sigmaQGR$ values, for the deviation parameters $\delta\chi_0$, $\delta\chi_4$, $\delta b_2$, and $\delta c_2$, obtained by varying one deviation parameter at a time.
Diamond markers denote TIGER analyses, while circles correspond to FTI results.
Since FTI probes only the inspiral phase of the waveform, results are shown only for $\delta\chi_0$ and $\delta\chi_4$ in that framework.
Each marker represents an individual simulated signal, with the color denoting the lensing regime to which it belongs.

A consistent pattern is observed across all deviation parameters.
Simulated signals in the geometric-optics (blue markers) and long-wavelength regimes (red markers) remain broadly consistent with GR, i.e., $\sigmaQGR \lesssim 1$.
In contrast, all deviations exceeding $3\sigmaQGR$ originate from the wave-optics regime (green markers), with several reaching the maximum measurable significance of approximately $4\sigmaQGR$.
This demonstrates that wave-optics lensing can be misinterpreted as a significant departure from GR when lensing effects are neglected in the recovery model.

Not every wave-optics dominated injection exhibits a large bias.
This behavior can be understood from the dependence of the amplification factor on both the impact parameter and the characteristic microlensing frequency.
For sufficiently large impact parameters ($y\gtrsim2$), the resulting interference pattern becomes weak, reducing the observable lensing signature.
Furthermore, the effectiveness of microlensing depends on where the induced modulations appear, relative to the signal power spectrum.
As discussed in Appendix~\ref{app:fml300}, GW150914-like signals become largely insensitive to microlensing once the characteristic modulation scale exceeds approximately $f_{\rm ML}\simeq300,{\rm Hz}$, because the underlying waveform amplitude has already decreased substantially at those frequencies.
Consequently, only a subset of wave-optics events produces distortions that are both sufficiently strong and located within the most sensitive portion of the detector band, leading to large apparent deviations from GR.

\paragraph*{\textbf{Modified Dispersion Tests:}}

\begin{figure}[htbp]
    \includegraphics[width=0.95\linewidth]{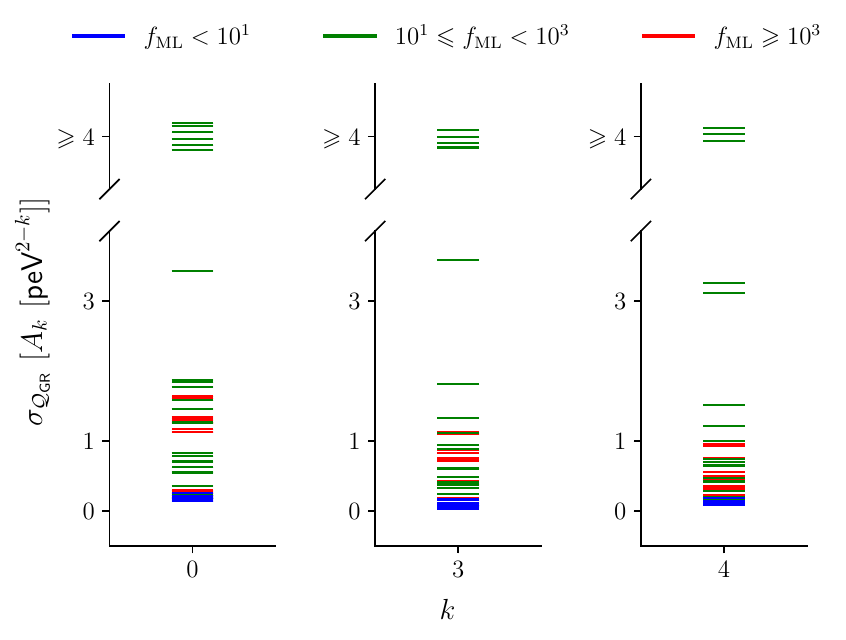}
    \caption{\label{fig:MDR}%
    Bias in MDR deviation parameters $A_k$, $k\in\{0,3,4\}$ for 50 GW150914-like microlensed injections. The figure follows the same format as Fig.~\ref{fig:PAR}. The y-axis shows the GR-deviation significance $\sigmaQGR$ obtained from the posterior distribution of the corresponding MDR parameter. Marker colors denote the microlensing regime of each simulated signal: long-wavelength (red), wave-optics (green), and geometric-optics (blue). We find that only wave-optics simulated signals exhibit deviations above $3\sigmaQGR$, while the remaining injections remain below $2\sigmaQGR$.}
\end{figure}

We next consider the MDR test, which probes deviations from non-dispersive propagation of GWs in vacuum.
Fig.~\ref{fig:MDR} summarizes the GR-deviation significance $\sigmaQGR$ for all $50$ simulated signals, obtained by constraining the MDR parameters $A_k$ with $k\in\{0,3,4\}$.
The three panels correspond to the three values of $k$, while the marker colors indicate the microlensing regime of the corresponding simulated signal.
The overall behavior closely mirrors that observed for the parameterized tests in Fig.~\ref{fig:PAR}.
Simulated signals belonging to the geometric-optics (blue) and long-wavelength (red) regimes remain consistent with GR, typically exhibiting $\sigmaQGR\lesssim1$ and never exceeding $2\sigmaQGR$.
In contrast, all statistically significant deviations arise from simulated signals in the wave-optics regime (green), with several events reaching the maximum measurable significance of approximately $4\sigmaQGR$.
As in the parameterized tests, not every simulated signal within the wave-optics regime exhibits a large deviation.

\paragraph*{\textbf{Consistency Tests:}}

\begin{figure}[htbp]
    \includegraphics[width=\linewidth]{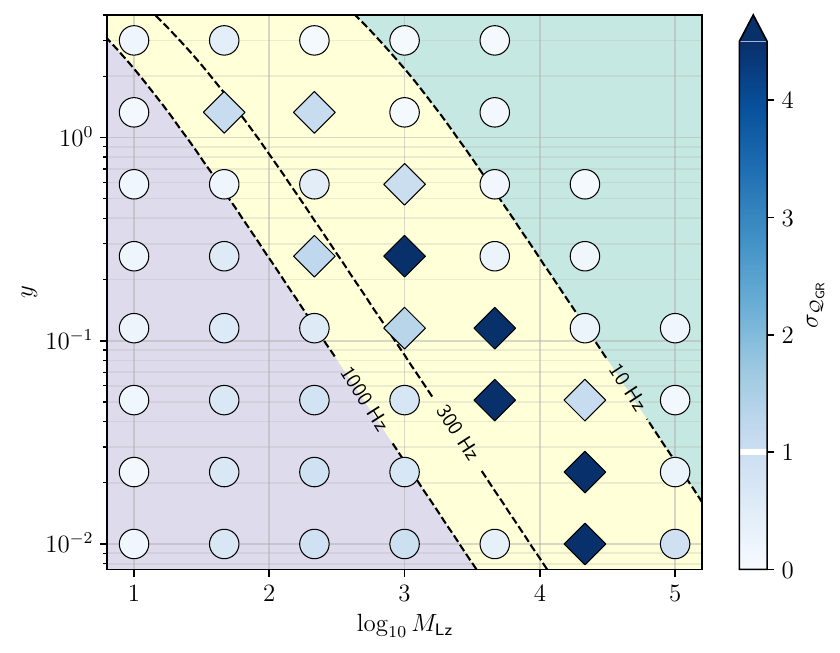}
    \caption{\label{fig:IMR}%
    IMRCT results for the 50 GW150914-like microlensed injections in the lens parameter space $\{\log_{10} M_{\rm Lz}, y\}$. Foreground markers are colored according to the GR-deviation significance $\sigmaQGR$, while the background follows the same pattern as in Fig.~\ref{fig:bayes}. Events with $\sigmaQGR\geq1$ are shown as diamonds. The missing marker at ($\log_{10} M_{\rm Lz}=3.66, y=0.023$) corresponds to a simulated signal for which the post-inspiral PE did not converge due to insufficient SNR. Significant deviations (diamond markers) are confined to the wave-optics regime and occur predominantly for $\fml\lesssim 300$ Hz.}
\end{figure}
\begin{figure}[htbp]
    \includegraphics[width=\linewidth]{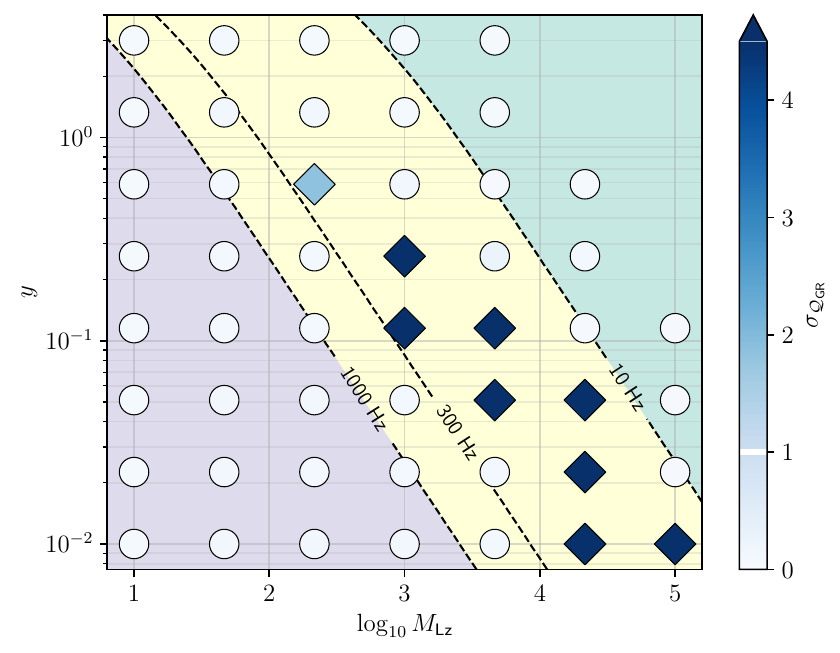}
    \caption{\label{fig:m-IMRCT}%
    Same as Fig.~\ref{fig:IMR}, but for meta-IMRCT. The overall trend remains unchanged, with the largest apparent GR violations occurring in the wave-optics regime. Relative to IMRCT, some deviations are enhanced while others are suppressed, indicating that combining multiple tests can either reinforce or mitigate microlensing-induced inconsistencies.
    }
\end{figure}
We finally consider the consistency tests of GR.
For each microlensed injection, a standard PE was first performed, assuming an unlensed model.
The MAP parameters from this analysis were then used to determine the inspiral--post-inspiral transition frequency through the corresponding $f_{\rm ISCO}$.
The resulting values lie in the range $[128.85, 167.22]~{\rm Hz}$, with one outlier at $68.92~{\rm Hz}$%
\footnote{
    For this outlier injection, the recovery is strongly biased towards a larger chirp mass, although the posteriors remain well converged.
}.
To ensure that the consistency tests provide meaningful results, we verified that the post-inspiral part of the signal contains sufficient SNR.
The optimal post-inspiral network SNRs lie in the range $[6.74,21.84]$, a consequence of fixing all simulated signals to a network SNR of $30$.

Representative posterior distributions for the IMRCT fractional deviation parameters $(\Delta M_f / \bar M_f, \Delta \chi_f / \bar \chi_f)$ were already presented in Fig.~\ref{fig:posteriors}.
We summarize the behavior of all $50$ injections through the corresponding GR-deviation significance $\sigmaQGR$ defined in Eq.~\eqref{eq:sigma-QGR}.
Figure~\ref{fig:IMR} shows the IMRCT results across the microlens parameter space.
The background shading denotes the three microlensing regimes, while the foreground markers encode the value of $\sigmaQGR$, with diamonds indicating simulated signals exhibiting deviations larger than $1\sigma$.

A trend consistent with the previous null tests is observed.
Simulated signals in the long-wavelength and geometric-optics regimes remain largely consistent with GR.
Significant apparent deviations arise within the wave-optics dominated regime, where the largest IMRCT deviations predominantly occur for $\fml\lesssim300~{\rm Hz}$.
This threshold is consistent with the transition identified from the Bayes factor and FF analyses and reflects the fact that lensing-induced modulations become most relevant when they overlap with the frequency range carrying significant signal power (see Appendix~\ref{app:fml300}).

The corresponding meta-IMRCT results are shown in Fig.~\ref{fig:m-IMRCT}.
The overall picture remains unchanged: the largest apparent deviations from GR are confined to the wave-optics regime, while simulated signals in the long-wavelength and geometric-optics regimes remain consistent with GR.
This independent confirmation is particularly noteworthy because meta-IMRCT combines information from multiple null tests and therefore probes a different aspect of the inference than IMRCT alone.
Comparing Figs.~\ref{fig:IMR} and \ref{fig:m-IMRCT}, however, reveals that the meta-analysis does not simply increase the significance of all events.
Instead, some simulated signals exhibit larger $\sigmaQGR$ values, while others become less significant.
This behavior indicates that the individual GR tests respond differently to the microlensing-induced waveform distortions.
When the biases inferred from different tests are mutually consistent, the combined meta-IMRCT analysis enhances the apparent deviation from GR.
Conversely, when the biases point in different directions, the combined significance is reduced.
The meta-analysis therefore acts as a coherence test, preferentially amplifying robust microlensing-induced inconsistencies while suppressing less coherent ones.

\begin{figure*}[htb]
    \includegraphics[width=0.9\linewidth]{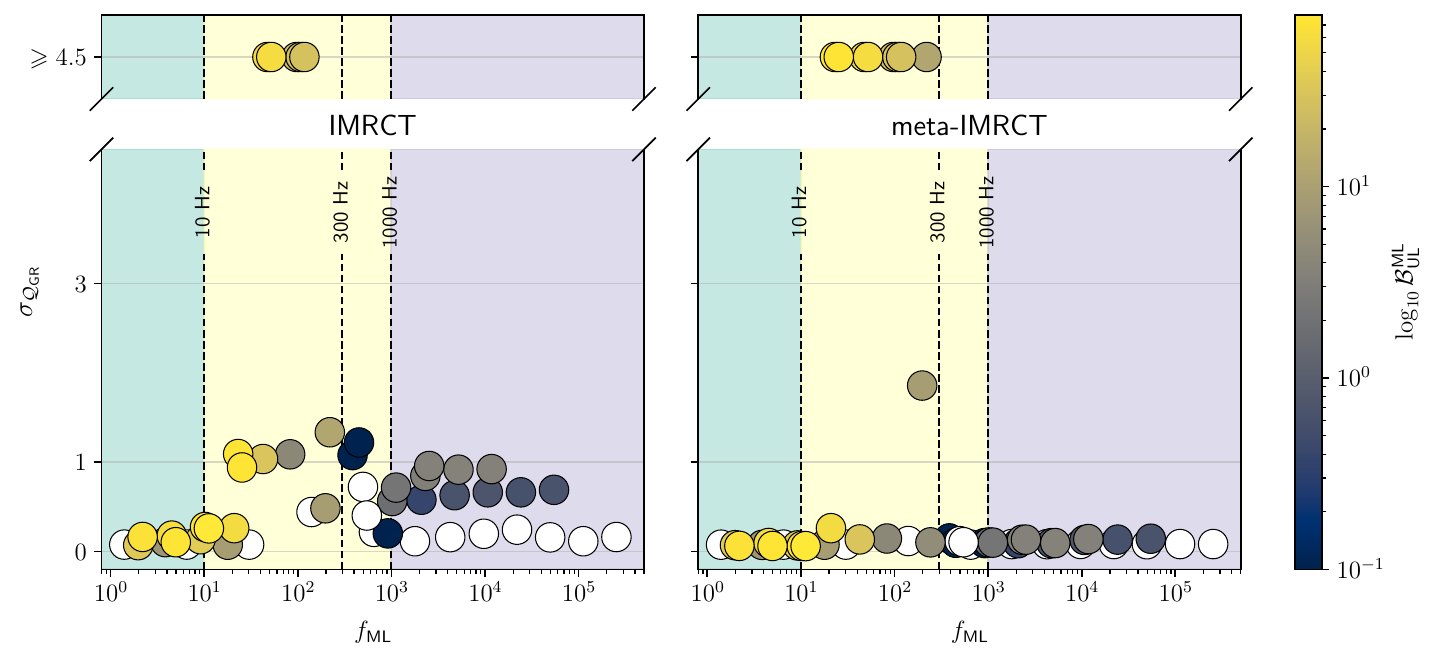}
    \caption{\label{fig:IMR-mIMR-fML-QGR-Bayes}%
    IMRCT (left) and meta-IMRCT (right) results shown as a function of the characteristic microlensing frequency $\fml$. The vertical axis denotes the GR-deviation significance, $\sigmaQGR$, while marker colors indicate the Bayes factor $\log_{10}\bayes$ in favor of the microlensed hypothesis over the unlensed hypothesis. White circles correspond to negative Bayes factors. The background shading denotes the long-wavelength, wave-optics, and geometric-optics regimes. We find that simulated signals with similar Bayes factors can exhibit substantially different GR-deviation significances, and vice versa.}
\end{figure*}
Figure~\ref{fig:IMR-mIMR-fML-QGR-Bayes} presents the IMRCT and meta-IMRCT results as a function of the characteristic microlensing frequency $\fml$, with marker colors indicating the Bayes factor $\log_{10}\bayes$.
This representation complements Figs.~\ref{fig:IMR} and \ref{fig:m-IMRCT} by combining the two-dimensional lens parameter space into a single physically relevant quantity and enabling a direct comparison between GR-deviation significance and microlensing detectability.
Consistent with our earlier observations, significant apparent deviations from GR are confined to the wave-optics regime and are not observed for $\fml \gtrsim 300~{\rm Hz}$.
The figure also illustrates the different behavior of IMRCT and meta-IMRCT.
While both methods identify the same region of parameter space as being susceptible to microlensing-induced biases, the meta-analysis tends to enhance the significance of the highly biased simulated signals while suppressing less significant ones.
Quantitatively, meta-IMRCT (IMRCT) yields 8 (5) simulated signals with deviations exceeding $3\sigmaQGR$, whereas 9 (11) simulated signals exceed $1\sigmaQGR$.

Finally, no simple correspondence is observed between the Bayes factor and the inferred GR-deviation significance.
Simulated signals with comparable values of $\log_{10}\bayes$ can exhibit substantially different values of $\sigmaQGR$, and vice versa.
Several injections at low $\fml$ strongly support the microlensed hypothesis while remaining largely consistent with GR.
The implications of this behavior are discussed in Sec.~\ref{sec:discussion}.

\section{Discussion \& Conclusions}
\label{sec:discussion}

In this work, we investigated the extent to which microlensing of GWs by compact objects biases a few of the standard tests of GR employed by the LVK collaboration.
These tests include parameterized tests (TIGER and FTI), propagation test (MDR), and consistency tests (IMRCT and meta-IMRCT).
Using a set of 50 GW150914-like microlensed injections covering the relevant lens parameter space, we quantified the degree to which diffractive lensing distortions are misinterpreted as apparent deviations from GR when analyzed with unlensed waveform models.

To enable a direct comparison between different tests, we introduced a unified measure of GR-deviation significance based on the GR quantile, $\mathcal{Q}_{\rm GR}$, and its corresponding Gaussian-equivalent significance, $\sigmaQGR$.
This framework provides a common statistical framework for both one-dimensional and multi-dimensional deviation posteriors and can be straightforwardly generalized to higher-dimensional tests of GR.
The principal conclusions of this work, reported in Sec.~\ref{sec:results}, can be summarized as follows:
\begin{itemize}
    \item \textbf{Microlensing-induced GR biases are confined to the wave-optics regime.}
    Significant apparent deviations from GR, reaching levels of approximately $4\sigma$--$4.5\sigma$, occur only when the lensing-induced waveform modulations overlaps with the frequency range carrying substantial signal power.
    For the GW150914-like simulated signals considered here, this occurs primarily for $\fml \lesssim 300~{\rm Hz}$.
    Simulated signals in the long-wavelength and geometric-optics regimes remain consistent with GR across all tests considered.
    \item \textbf{Microlensing detectability does not strongly correlate with biases of GR deviation parameters.}
    Some simulated signals with large $\log_{10}\bayes$ show weak or no apparent GR violation.
    \item \textbf{Unlensed baselines remain well behaved.}
    Rerunning all GR tests on an unlensed GW150914-like injection yields deviations $\lesssim0.7\sigma$, confirming that the large apparent deviations originate solely from the waveform systematics induced due microlensing effects.
\end{itemize}
Together, these results demonstrate that microlensing as an astrophysical source of waveform systematics that can bias precision tests of GR, particularly for future detectors that will probe higher redshifts and detect signals with higher SNRs.

A particularly important result of this study is the weak correlation between microlensing detectability and GR-test bias.
The Bayes factor quantifies the overall preference for a microlensing modulation relative to an unlensed one and is therefore sensitive to the total waveform mismatch accumulated across the detector bandwidth.
In contrast, each GR test is sensitive only to specific projections of the microlensing-induced amplitude and phase modulations that resemble its corresponding deviation parameter.
Consequently, a lensing-induced distortion can be highly detectable while projecting only weakly onto the deviation parameters of a given GR test.
Conversely, a comparatively modest lensing signature can closely mimic the frequency dependence of a particular deviation parameter and therefore produce a significant apparent GR violation.

The results presented here have several implications for future tests of gravity with GWs.
Propagation effects associated with gravitational microlensing should be regarded as an astrophysical source of waveform systematics and may therefore bias precision tests of GR if not properly accounted for.
The importance of these effects is expected to increase for next-generation observatories such as the Einstein Telescope~\cite{Punturo:2010zz} and Cosmic Explorer~\cite{Reitze:2019iox}, which will detect signals with significantly higher SNRs and probe source populations at substantially larger redshifts where the probability of lensing is higher.
Furthermore, population-level analyses that combine information from many events may accumulate small microlensing-induced biases in ways that could mimic weak departures from GR.

Perhaps most importantly, our results indicate that lensing detectability alone is not a sufficient safeguard against false GR violations.
Since the magnitude of the inferred bias does not correlate strongly with the Bayes factor for microlensing, simply excluding events with strong lensing support would not eliminate all potentially problematic cases.
Future precision tests of GR will therefore benefit from waveform models that incorporate wave-optics lensing effects, or from inference frameworks capable of marginalizing over lensing-induced waveform systematics, thereby ensuring that astrophysical propagation effects are not misinterpreted as evidence for new gravitational physics.

\acknowledgments
We thank Haris K. for reading the manuscript and comments.
AK would like to thank Sudhir Gholap for many fruitful discussions contributing to this work. AK’s research was supported by the University Grants Commission, Government of India.
NVK acknowledges support from STFC grant ST/Y00423X/1. AM acknowledges the support of the Department of Atomic Energy, Government of India, under project nos. RTI4019 and RTI4013.
The authors are grateful for computational resources provided by IUCAA, Pune, India, LIGO Laboratory and Leonard E Parker Center for Gravitation, Cosmology and Astrophysics at the University of Wisconsin-Milwaukee, which are supported by National Science Foundation Grants PHY-0757058, PHY-0823459, PHY-1626190, and PHY-2110594.

The work utilizes the following software packages:
\texttt{Cython}~\cite{cython},
\texttt{NumPy}~\cite{Harris:2020xlr},
\texttt{SciPy}~\cite{Virtanen:2019joe},
\texttt{astropy}~\cite{astropy},
\texttt{PyCBC}~\cite{pycbc},
\texttt{LALSuite}~\cite{lalsuite, swiglal},
\texttt{dynesty}~\cite{Speagle:2019ivv},
\texttt{BILBY}~\cite{Ashton:2018jfp, Romero-Shaw:2020owr},
\texttt{corner}~\cite{corner},
\texttt{Matplotlib}~\cite{Hunter:2007ouj}, and
\texttt{Jupyter notebook}~\cite{jupyter}.

\appendix

\section{Why does $\fml\simeq 300~{\rm Hz}$ mark the transition for GW150914-like simulated signal?}
\label{app:fml300}

Throughout Sec.~\ref{sec:results}, we found that the strongest microlensing signatures occur predominantly for $\fml \lesssim 300~{\rm Hz}$.
Above this value, both the microlensing detectability and the inferred GR biases rapidly diminish.
In this appendix, we explain the physical origin of this transition for a GW150914-like binary.
The key point is that microlensing-induced waveform modulations become observable only when they occur within the frequency range carrying significant signal power.
Even if a signal formally lies in the wave-optics regime, the resulting distortions will have little impact on parameter estimation or GR tests if they appear primarily at frequencies where the GW amplitude is already strongly suppressed.

\begin{figure}[htbp]
    \includegraphics[width=\linewidth]{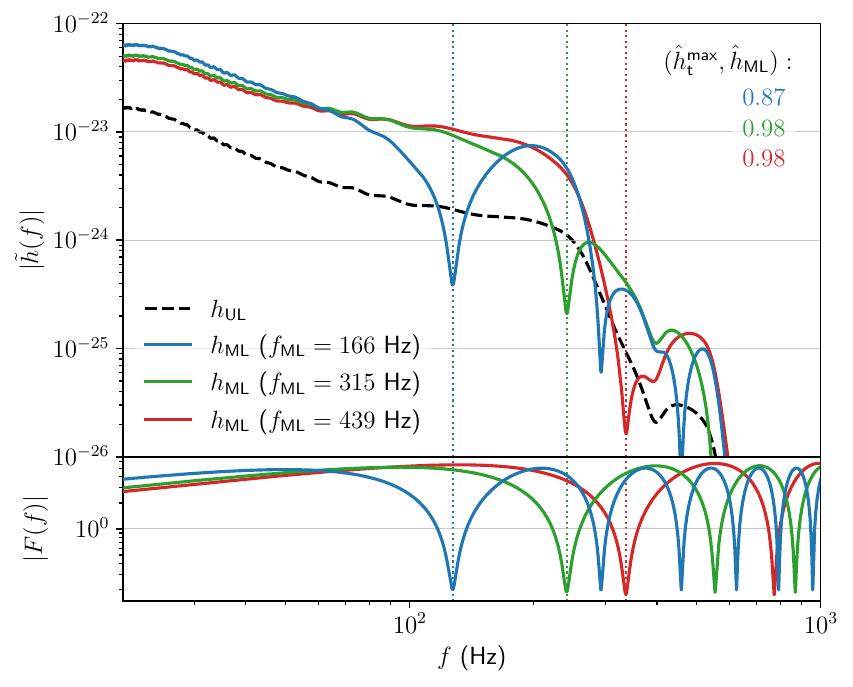}
    \caption{\label{fig:fml300}
    GW150914-like unlensed ($h_{\rm UL}$) and microlensed ($h_{\rm ML}$) signals (top panel), together with the microlensing amplification factor $F(f)$ (bottom panel). Three representative simulated microlensed signals are shown with $\fml$ below (blue), near (green), and above (red) $300~{\rm Hz}$. The match between these injections and their corresponding best-fit unlensed template ($h_t^\text{max}$) from FF analysis is indicated in the top-right corner (colored according to the legend). Vertical dotted lines mark the frequencies at which the first significant modulation appears for different microlensing configurations (i.e., different $\fml$ values).
    The amplitude of the unlensed waveform decreases by approximately one order of magnitude near the green vertical line, implying that lensing modulations appearing at or beyond this frequency contribute weakly to the accumulated SNR. Consequently, the $\fml\lesssim300~{\rm Hz}$ case exhibits a substantially larger mismatch ($\sim13\%$) than the two cases with $\fml\gtrsim300~{\rm Hz}$, which is $\sim2\%$.
    }
\end{figure}

To illustrate this behavior, we consider three representative points in the microlens parameter space, $(M_{\rm Lz}, y) \approx [(2\times 10^3,0.08),~(1.2 \times 10^3, 0.07),~(10^3,0.06)]$, corresponding to $\fml=166$, $315$ and $439~{\rm Hz}$, respectively, i.e., below, near and above the transition frequency inferred from Fig.~\ref{fig:bayes}.
Figure~\ref{fig:fml300} shows the corresponding GW150914-like simulated microlensed and unlensed signals.
The bottom panel, which shows the corresponding lensing amplification factor $F(f)$, demonstrates that the characteristic frequency scale of the oscillatory lensing pattern is determined by $\fml$.
As $\fml$ increases, the first prominent modulation shifts toward higher frequencies.
For the GW150914-like signal considered here, the waveform amplitude has already decreased by approximately an order of magnitude by the time the modulation reaches frequencies near $300~{\rm Hz}$.
Consequently, lensing-induced distortions appearing at or above this frequency contribute only weakly to the matched-filter SNR and therefore have little impact on parameter estimation.

This behavior is also reflected in the FF analysis.
The $\fml=166~{\rm Hz}$ injection exhibits pronounced waveform distortions across the most sensitive part of the detector band, producing a mismatch of approximately $13\%$ with the best-fit unlensed template.
In contrast, the $\fml=315~{\rm Hz}$ and $\fml=439~{\rm Hz}$ injections differ from the unlensed waveform only in the high-frequency portion of the signal, resulting in mismatches of only ${\sim}2\%$.
Since the Bayes factor in favor of microlensing is ultimately driven by the mismatch between microlensed and unlensed waveforms, its value decreases rapidly once $\fml$ exceeds its threshold.
This behavior is precisely what is observed in Fig.~\ref{fig:bayes}.
The same argument also explains the behavior of the GR tests presented in Sec.~\ref{sec:results}.
Apparent deviations from GR arise only when microlensing-induced phase and amplitude distortions are sufficiently strong to bias the recovery.
Once the characteristic lensing modulation moves beyond the frequency range carrying significant signal power, the recovered parameters become largely insensitive to the microlensing effects.
As a result, the strongest apparent GR violations are confined to the same region, namely $\fml\lesssim300~{\rm Hz}$.

It is important to emphasize that the value $300~{\rm Hz}$ is not universal.
The frequency at which the unlensed signal loses most of its power depends on both the source properties and the waveform approximant used for recovery.
For inspiral-only approximants, the relevant scale is typically the ISCO frequency, $f_{\rm ISCO}\propto M_{\rm tot}^{-1}$, and the corresponding transition in $\fml$ is expected to be of the same order.
Waveform models that include merger and ringdown emission extend to higher frequencies and therefore remain sensitive to lensing-induced modulations over a broader frequency range.
For the GW150914-like system analyzed here using \texttt{IMRPhenomXPHM}, this transition occurs near $300~{\rm Hz}$, which is larger than the corresponding ISCO frequency because appreciable signal power remains in the merger-ringdown regime.

This raises a natural question: if the practical transition for GW150914-like signals occurs near $300~{\rm Hz}$, why do we define the wave-optics regime more broadly as $10~{\rm Hz} < \fml \leq 1000~{\rm Hz}$?
The reason is that our goal is to characterize microlensing-induced biases independent of any specific compact binary.
The frequency at which lensing effects become observationally relevant depends on the intrinsic source parameters, particularly the total mass.
Since detectable compact binaries can have characteristic frequencies spanning much of the LVK observing band, a source-independent classification should not be tied to a single value such as $300~{\rm Hz}$.
We therefore adopt the broader interval $10~{\rm Hz}<\fml\leq1000~{\rm Hz}$ as the wave-optics regime, while explicitly highlighting the $\sim300~{\rm Hz}$ transition wherever relevant for the GW150914-like injections considered in this work.

\bibliography{bibliography}

\end{document}